\documentclass[a4paper]{cas-sc}

\usepackage{array}
\usepackage{graphicx}
\usepackage[dvipsnames]{xcolor}
\usepackage{url}
\usepackage{amsmath, amsthm, amssymb}
\usepackage{booktabs}
\usepackage[authoryear,longnamesfirst]{natbib}
\usepackage[normalem]{ulem}

\usepackage[center, font=footnotesize]{caption}
\usepackage[caption=false,font=footnotesize]{subfig}

\usepackage{makecell}

\usepackage{tabularx}

\newcommand{\tabitem}{~~\llap{\textbullet}~~}

{\ignorespacesafterend}

\newcommand{\ie}{\textit{i.e.}, }
\newcommand{\eg}{\textit{e.g.}, }

\graphicspath{{figures/}}

\begin{document}

\title[mode = title]{Network-wide assessment of ATM mechanisms using an agent-based model}

\shorttitle{Network-wide assessment of ATM mechanisms}
\shortauthors{Delgado, Gurtner,  Mazzarisi, Zaoli, Valput, Cook and Lillo}
\shortauthors{}

 \author[1] {Luis Delgado}[orcid=0000-0003-4613-4277]
 \fnmark[1] 
 \cormark[1]

 \author[1]{Gérald Gurtner}[orcid=0000-0003-2006-4653]
 \fnmark[1]

 \author[2]{Piero Mazzarisi}[orcid=0000-0002-7198-513X]
 \fnmark[1]

\author[2] {Silvia Zaoli}[orcid=0000-0001-6619-0984]
 \fnmark[1] 

 \author[3] {Damir Valput}[orcid=0000-0002-0797-1956]

 \author[1] {Andrew Cook}

 \author[2] {Fabrizio Lillo}[orcid=0000-0002-4931-4057]

 \address[1]{School of Architecture and Cities, University of Westminster, London, United Kingdom}
\address[2]{Dipartimento di Matematica, University of Bologna, Bologna, Italy} 
 \address[3]{Innaxis, Madrid, Spain}

 \cortext[cor1]{Corresponding author} 
 \fntext[fn1]{These authors have contributed equally to the present work.}

\date{\today}
\begin{NoHyper}
\maketitle
\end{NoHyper}

\begin{abstract}
This paper presents results from the SESAR ER3 Domino project. Three mechanisms are assessed at the ECAC-wide level: 4D trajectory adjustments (a combination of actively waiting for connecting passengers and dynamic cost indexing), flight prioritisation (enabling ATFM slot swapping at arrival regulations), and flight arrival coordination (where flights are sequenced in extended arrival managers based on an advanced cost-driven optimisation). 
Classical and new metrics, designed to capture network effects, are used to analyse the results of a micro-level agent-based model. 
A scenario with congestion at three hubs is used to assess the 4D trajectory adjustment and the flight prioritisation mechanisms. Two different scopes for the extended arrival manager are modelled to analyse the impact of the flight arrival coordination mechanism.
Results show that the 4D trajectory adjustments mechanism succeeds in reducing costs and delays for connecting passengers. A trade-off between the interests of the airlines in reducing costs and those of non-connecting passengers emerges, although passengers benefit overall from the mechanism. Flight prioritisation is found to have no significant effects at the network level, as it is applied to a small number of flights. Advanced flight arrival coordination, as implemented, increases delays and costs in the system. The arrival manager optimises the arrival sequence of all flights within its scope but does not consider flight uncertainties, thus leading to sub-optimal actions.

\end{abstract}

\section{Introduction}
\label{sec:intro}

Understanding the complex interdependencies and coupling of various components of the ATM system is a major challenge for ATM architects. Assessing how the introduction of changes in one (sub)system may impact others, not only locally, but also at the network level, is particularly difficult. The `Airspace Architecture Study’ \citep{SJU_I} highlights how the ATM system can be understood as nodes, which are often operating close to maximum capacity and without appropriately connected resources: disruptions in the system often propagate and "knock it out of optimal flow”. This ‘domino’ effect is offset, for example, by strategically placed buffers in schedules, although these are often insufficient to absorb all of the disturbance(s). The Study indicates the need for “stronger linking between airspace, operations and technical evolution and measurement of the impact through simulations factoring in known deployments ...”. It also recommends targeted incentives for “early movers”. The `Transition Plan' \citep{SJU_II} further endorses the acceleration of market uptake of the next generation SESAR technologies and services to support defragmentation. The Plan identifies three key operational and technical measures that need to be implemented in the very short term (2020 to 2025), to initiate the changes outlined. These measures will allow different parts of the system to be developed at different paces, while maintaining coherence at the network level and awareness of local needs.

Previous research has highlighted the need to consider different stakeholders in order to fully understand the impact of introducing changes in the ATM system \citep{COOK201638, vista}. Domino, an H2020 -- SESAR JU Research and Innovation project, expands some of these concepts by further developing a model (Mercury) to assess the coupling of ATM systems from a flight and passenger perspective. Mercury is able to model a whole day of operations at the ECAC level following flights and passengers \citep{domino53}. It allows ATM system designers to better understand the relationships between (sub)systems and the nature of such relationships, which emerge in a given technological and operational context. The outcome of the model is analysed with the use of classical and advanced network metrics (centrality and causality), which afford new insights into the impact of modifying the ATM system at the network level – see \citep{zaninlillo,cook_complexity_atm} for an early review of the value of non-classical metrics and the need to differentiate between flight- and passenger-centric indicators.

Exploring several issues outlined in the Airspace Architecture Study \citep{SJU_I}, in the context of different (future) operational and stressed environments, this paper presents the assessment of three mechanisms: 4D flight trajectory adjustments (4DTA), flight prioritisation (FP) and flight arrival coordination (FAC). These mechanisms are modelled considering a baseline, which resembles current practices, and as an advanced implementation, in two distinct scenarios. \\
4DTA is composed of two sub-mechanisms: (i) dynamic cost indexing (DCI), whereby a flight is able to adjust its cruise speed to manage expected delay \citep{dci}; and (ii) waiting for passengers (WfP), whereby a flight can wait at-gate for late connecting passengers \citep{cass_II, COOK201638}. FP is inspired by UDPP concepts allowing the swapping of ATFM slots among airlines \citep{UDPP_SID_2016,UDPP_PJ7}. FAC provides advanced management of fight arrivals for the sequencing and merging thereof \citep{eaman_wkshp,Montlaur2017}.

The paper is organised as follows. Section~\ref{sec:model} introduces the background concerning the model developed in Domino (Mercury), as well as its main characteristics. It also provides a literature review on the use of agent-based models (ABM) in ATM and the contributions of Mercury towards the development of advanced tools for the assessment of air traffic management initiatives. Section~\ref{sec:mechanisms} describes the three mechanisms analysed in this paper in more detail and outlines the advancement on the state of the art with respect to these mechanisms. The outcome of the model needs to be analysed in order to gain an insight into the impact of the mechanisms in the system. In particular, dedicated advanced network metrics have been developed as part of the Domino project. These metrics are described in Section~\ref{sec:indicators}. Two scenarios are presented in Section~\ref{sec:scenarios} and the main findings summarised in Section~\ref{sec:results}. The paper closes with conclusions and proposals for future work, in Section~\ref{sec:conclusions}.

\section{Model -- Mercury}  
\label{sec:model}

\subsection{Agent-based models and air traffic management}

Agent-based models (ABM) used as the basis for a network approach of system design are relatively new in the ATM domain. ABMs that can be found in this field in the literature \citep{Bouarfa2013, Stroeve2013, Molina2014, Gurtner2017, Delgado2017} are usually used to assess the impact of fairly limited changes in the airspace and/or are focused on a relatively small geographical area. For instance, some research projects modelled passenger flows in airport terminals as multi-agent systems \citep{enciso2016, schultz2011}. Other studies focused on modelling the behaviour of individual and groups of passengers and their impact on passenger flows \citep{cheng2014, chen2018}.

Models encompassing all of the European traffic at once are rare and usually confidential in terms of documentation and implementation (for instance EUROCONTROL's internal tool RNEST\footnote{\url{https://www.eurocontrol.int/simulations} (accessed February 2021)}). Moreover, these models are often deterministic, missing a crucial ingredient of the airspace -- its variability -- and/or include very simple rule-based behaviours for their agents. Finally, to the authors' knowledge, almost none of these tools are able to compute passenger metrics.

Mercury was developed in response to that need, over the course of several years \citep{POEM, COOK201638, vista}. This allowed one of the first European-wide passenger KPI assessments in terms of travelling times, and was later expanded to include door-to-door travel times \citep{Kluge_18}. The Domino model is a new expanded version, the main features of which have been presented in \citep{Mazzarisi2018}.

\subsection{Model description}

Mercury is an ECAC-wide micro-level agent-based model (ABM) comprising the following main agents:
\begin{itemize}
    \item `Airline operating centre' (AOC): it is responsible for the management of the airline's fleet (managing the dispatching, cancellation, ATFM slot swapping, etc. of the flights), and of the passengers (rebooking passengers when missed connections occur, compensating them, etc.). The AOC decisions are based on the estimation of the expected cost of delay, calibrated with \citep{cost_delay}. It is to all effects the most complex agent.
    \item `Flight': there is a flight agent per individual flight in the simulation. These agents integrate the trajectories based on the flight plans. They also performs basic flight operations such as requesting for a departing slot.
    \item `Ground airport': it captures and provides estimation and realisation times of the different on-ground airport processes: taxi-in and taxi-out, turnaround times, and passengers connecting times.
    \item `Network Manager': it accepts or rejects flight plans, issues ATFM delays from probabilistic or explicitly modelled regulations.
    \item `AMAN': it manages the arrival sequence of flights to ensure that airport arrival capacities are respected. An airport might implement an extended arrival manager (E-AMAN), with an extended planning and an execution horizon, or just a simpler arrival manager (AMAN). With the E-AMAN, when a flight reaches the planning horizon, a slot is pre-assigned enabling the possibility of adjusting their trajectory (reducing their cruising speed) to save fuel while performing part of the required delay. 
    Once flights enter the execution horizon, they are finally sequenced and holding delays issued. A first come first served approach is used in airports with a simple AMAN.
    \item `DMAN': this type of agents provides slots to flights following a first come first served policy, assigning flights to a queue of departing slots to ensure that the runway capacities are respected. The delay issued will be performed at the gate prior push-back. 
    \item `Radar': this singleton agent follows the flights trajectories and notifies interested parties (e.g., AMAN, AOC) when significant milestones are reached (e.g., top of climb (TOC), entering the planning horizon).
    \item `Flight swapper': The flight swapper is responsible for the management of ATFM slots swaps following the principles of User-Driven Prioritisation Process (UDPP) \citep{UDPP_SID_2016, UDPP_PJ7}.
\end{itemize}

There is one instance of the `Airline' agent per individual airline and one `Flight' agent per flight. Passengers typically have only few decisions to make and thus are not modelled as agents, even though their preferences are taken into account by the airlines indirectly through a soft cost based on a logit utility function (calibrated again with \citep{cost_delay}). Each airport has an instance of the `Ground airport', `DMAN' and `AMAN' agents. Finally, the simulation counts with a single instance of the `Network Manager', `Radar' and `Flight swapper' agents.

Mercury is executed as an event-driven simulator. In addition to events (\eg `flight pushback'), agents also react to messages from other agents (\eg a request for rebooking passengers which have missed connections). Random variables are used for uncertainties (\eg taxi times, cruise wind, passengers connecting times, turnaround times, and cancellations), and for some non-optimal decision-making processes by agents (\eg as part of the dispatch process the selecting the flight plan to be operated). All distributions are built on various empirical data and the model has been calibrated with historical data (see Section~\ref{subsec:calibration}). Due to the stochasticity, for each scenario several runs of the model are performed.

In the model, the agents (particularly the AOC) estimate their expected costs and try to perform actions to minimise them. 
In particular the model considers the following costs:
\begin{itemize}
    \item `cost of delay', estimated as the extra cost that airlines experience on their flights due to delay. These costs can be broken down and examined for different types of aircraft and companies, as in \citep{cost_delay}. In particular, Mercury considers:
    \begin{itemize}
        \item `non-passenger related' costs: extra cost on maintenance and crew due to delay, based on \citep{cost_delay}.
        \item `passenger related' costs, which are divided between:
        \begin{itemize}
            \item `hard' costs, which have a direct monetary translation, such as compensation costs for passengers due to Regulation 261 \citep{reg261}, or cost associated with duty of care, re-booking of missed connections, etc. These costs are modelled explicitly in Mercury.
            \item `soft' costs, \ie costs that represent a future loss for the company in terms of market share due to deteriorated perception by the passenger. The estimation of these costs is historically based.
        \end{itemize}
    \end{itemize}
    \item cost due to breaching curfews: flights breaching curfews may incur a wide range of costs, from a small fine to an interdiction to land.
    Mercury implements `hard' curfew: if a flight plan is expected to arrive after the curfew threshold, it is rejected by the Network Manager. This protocol is implemented in some selected airports. A `curfew buffer' is computed and dynamically updated for each flight, \ie the maximum delay that a flight can experience before risking generating reactionary delay which might end up in a curfew breach. This might impact the behaviour of the AOC, \eg recovering early morning delays to avoid later curfews.
    \item fuel costs: Mercury models the nominal fuel usage and its associated cost (0.5 EUR/kg), but also considering tactical uncertainties, \eg wind, and the impact on fuel of delay management strategies, \eg holdings or adjustments to cost index. Fuel usage is estimated using BADA 4 performances~\citep{eurocontrol_bada}.
    \item airspace charges: they are computed considering the 39 regions managed by EUROCONTROL CRCO plus the surrounding airspaces of other countries with different charging schemes. These costs allow the AOC to select which flight plan to submit to the Network Manager comparing the total expected operational costs of different alternatives routes even when they use adjacent airspaces to the European core.
\end{itemize}

\noindent
It is worth noting that the cost of delay is typically a non-linear function of delay. Negative delays (flights that arrive before schedule) are usually of no monetary value to the airline. High delays are in proportion more costly than small delays (cost increase on average as roughly quadratic with delay \citep{cost_delay}). Finally, the cost functions tend to exhibit significant noncontinuous increments linked to events such as passengers missing connections.

Mercury produces detailed metrics for each flight and passenger in the system, \eg departing, arrival delays, taxi times, missed connections, compensations. Aggregated values (KPIs) are then computed based on this output.

\subsection{Mercury inputs}

Mercury requires a significant amount of data to define all the parameters needed to execute a simulation. The input consists of:
\begin{itemize}
    \item data required to model airline operations and flights
    \begin{itemize}
        \item flight schedules,
        \item pre-computed flight plans covering, for each aircraft type, the different possible routes between origins and destinations; used by the AOC to select operational flights plan for their flights,
        \item data on aircraft performance models (BADA 4),
        \item general airlines information (e.g., alliances)
    \end{itemize}
    \item passenger itineraries,
    \item ATM operational data
    \begin{itemize}
        \item probabilities of ATFM delay per ANSP for ATFM regulations issued in the airspace,
        \item samples of ATFM regulations issued at airports,
        \item distributions of turnaround times, connecting times, taxi-times, non-ATFM-delay for each airport,
        \item airport data, such as runway capacity, curfews or usage of arrival managers,
        \item cost of delay parameters
    \end{itemize}
    \item other operational parameters, such as the cost of fuel or en-route charges,
    \item hyperparameters required to define scenario parameters, such as for each mechanism their implementation level (baseline, advanced)
\end{itemize}

\subsection{Datasets, calibration and validation}
\label{subsec:calibration}

Agent-based models need to be calibrated to produce reliable results. Even if the models are able to simulate any particular input, the quality of the results depends on the calibration of some distributional parameters. To calibrate Mercury we used various data sources, summarised in Table~\ref{tab:data_sources}, along with consultation with relevant stakeholders (e.g., airlines) to capture their behaviour. 

\begin{table}
    \centering
    \begin{tabularx}{\textwidth}{c|X|X}
        Data source & Main usage & Reference \\
        \hline
        \hline
        DDR2 & Used to get the set of flights, origin-destination, routes, aircraft type, estimated cruise wind, distributions on climb and descent profiles, requested nominal cruise speeds and flight levels, airlines, alliances, airspace structure, ATFM regulations & \citep{EUROCONTROL_DDR2}\\
        Cost of delay reporting & Used to compute cost of delay function &  \citep{cost_delay}\\
        IATA Summer Season 2010 from CODA & Taxi times & \citep{cassii}\\
        DDR2 & Minimum turnaround times, minimum connecting times & \citep{ComplexityCost}\\
        CODA & Non-ATFM delays & \citep{coda}\\
        Paxis, GDS & For passenger itineraries, including fares and classes & \citep{ComplexityCost}\\
        Innovata (Cirium) & Flight schedules & -- \\
    \end{tabularx}

    \caption{List of data sources used.}
    \label{tab:data_sources}
\end{table}

The datasets are restricted only to commercial flights landing or taking-off in Europe. DDR data are used to obtain flights characteristics relevant for the model, including possible routes and aircraft rotations, and to characterise other operational parameters, such as ATFM regulations, which are based on the analysis of historical DDR data for several AIRACs. Schedule data are used to extract the SOBT and SIBT of these flights. Passenger data are used to define individual itineraries and merge them with the flight data. Other data sources, e.g., CODA datasets, are used to calibrate certain parameters in the model as described below. 

Many processes in Mercury are modelled using statistical distributions. For example, the average wind encountered by flights during the en-route phase, or the amount of ATFM delay assigned to a flight due to an en-route regulation. Table~\ref{tab:processes} presents some of the key processes that are modelled in Mercury, and how the distributions of the associated indicators/parameters have been calibrated.

\begin{table}
    \centering    
    \begin{tabularx}{1\textwidth}{>{\hsize=.5\hsize}X|>{\hsize=1.5\hsize}X|X}
        Process & Distribution & Based on \\
        \hline
        \hline
        Taxi-in/out& Log-Normal distribution with modified
            mean, standard deviation for different scenarios & IATA Summer Season 2010 from CODA \\
        Climb uncertainty & Normal distribution, minutes & Analysis of DDR difference between planned and executed trajectories (M2, M3) from the DCI4HD2D project \citep{cassii} \\
        Cruise & Normal distribution, NM & Analysis of DDR difference between planned and executed trajectories (M2, M3) from the DCI4HD2D project \citep{cassii}\\
        Wind & Empirical probability distribution for planned wind during the cruise. Used percentile of wind between regions. No noise added on execution. & For each ANSP to ANSP origin and destination airport consider the difference between requested speed and observed average ground speed for cruise segments from DDR2 analysis (AIRAC1409).\\
        Turnaround time & Exponential distribution based on minimum turnaround time based on airport size, aircraft wake and type of airline. Distributions modified based on scenario & Analysis of turnaround times performed in the POEM project and used in the ComplexityCosts project \citep{ComplexityCost}\\
        ATFM delay & Airport regulations are sampled from a historical day. The day is selected based on their percentile ranked by the number of regulations at the airport on the day. Airspace regulation delays are based on two distributions, one for weather and one all other types of regulations & Based on analysis of DDR2 (AIRAC1313-1413 excluding days with industrial actions)\\
        ATFM delay & Empirical probability distribution function for regulations due to weather and regulations for other reasons. & Based on analysis of DDR2 (AIRAC1313-1413 excluding days with industrial actions)\\
        Non-ATFM delay & Exponential distribution with parameters modified based on scenarios & - \\
        Connecting times & Log-Normal distribution based on minimum connecting times per airport and type of connection (national-national, international-international and national-international) & Based on analysis of minimum connecting times at ECAC airports originally performed in the POEM project \citep{POEM}\\
        Variation of cruise length due to DCI & Normal distribution, NM & Analysis of performance using Airbus Performance Engineering Programme \citep{cassii}
    \end{tabularx}
    \caption{Calibration parameters in the model.}
    \label{tab:processes}
\end{table}

The calibration and validation has been performed by considering not only average values but the distribution of key indicators. For example, we analysed the distribution of arrival and departure delay as reported in DDR and CODA with the model~\citep{coda}.

\section{Mechanisms}
\label{sec:mechanisms}

\begin{table}
\centering
 \begin{tabularx}{1\textwidth}{|
    >{\hsize=0.55\hsize\linewidth=\hsize}X|
    >{\hsize=0.6\hsize\linewidth=\hsize}X|
    >{\hsize=0.6\hsize\linewidth=\hsize}X|
    >{\hsize=2.25\hsize\linewidth=\hsize}X|}
 \hline
 Mechanism & Rationale & Implementation & Characteristics \\ 
 \hline\hline
\multirow{7}{*}{\parbox{2.25cm}{4D Trajectory \newline Adjustments  (4DTA)}} & \multirow{7}{*}{\parbox{2.25cm}{Actively waiting for connecting passengers prior departure and adjusting cost index to manage delay.}} & \multirow{4}{*}{Baseline} & \tabitem Rule of thumb. \\ 
& & & \tabitem Dynamic Cost Index (DCI) prior departure, recovery probability as function of amount of departing delay. \\ 
& & & \tabitem Waiting for passengers (WfP) for flexible ticket holders up to 15 minutes. \\
& & & \tabitem Decoupled decision between DCI and WfP. \\\cmidrule{3-4}
& & \multirow{3}{*}{Advanced} & \tabitem DCI coupled with WfP prior departure.\\
& & & \tabitem Cost based decision. \\ 
& & & \tabitem Re-assessment of DCI decision at TOC. \\ 
 \hline
 
 \multirow{4}{*}{\parbox{2.25cm}{Flight \newline Prioritisation (FP)}} & 
 \multirow{4}{*}{\parbox{2.25cm}{Swapping of ATFM slots at arrival regulations.}} & Baseline & \tabitem No exchange of ATFM slots available. \\ \cmidrule{3-4}
& & \multirow{2}{*}{Advanced} & \tabitem Possible to swap slots for flights in same ATFM regulation at arrival airport.\\
& & &  \tabitem Possible intra- and inter- airlines slots swaps. \\ 
 \hline
 
  \multirow{4}{*}{\parbox{2.25cm}{Flight Arrival \newline Coordination (FAC)}} & 
  \multirow{4}{*}{\parbox{2.25cm}{Extended Arrival Manager at airports.}} & \multirow{2}{*}{Baseline} & \tabitem Assignment of arrival slots to first available slot. \\ 
 & & & \tabitem No AOC information used. \\ \cmidrule{3-4}
& & \multirow{2}{*}{Advanced} & \tabitem Request cost information from flight and AOC.\\
& & & \tabitem Assign slots which minimise expected costs. \\ 

 \hline
\end{tabularx}
\caption{Summary of mechanisms characteristics.}
\label{table:mechanisms}
\end{table}

Table~\ref{table:mechanisms} summarises the three mechanisms evaluated. These have been selected to cover different stakeholders (airlines, airports, network manager) and to have a spread on the location of their usage (localised at a given region, \eg at an airport, or across the network, \eg by AOCs on all flights). The detailed description of each mechanism is presented in the following sub-sections. Each mechanism has been executed in a baseline and and advanced implementation. 

The baseline implementation is based on current (2014) equivalent operational concepts, the advanced implementation is inspired, if more exploratory, by future aspects of SESAR~\citep{master_plan2015}. More information on the model details can be found in \citep{Mazzarisi2018,domino41}

\subsection{4D trajectory adjustments (4DTA)}
\label{subsec:4DTA_desc}

4D trajectory adjustments is a mechanism which modifies the flights trajectories to deal with delay. This mechanism is formed of two sub-mechanisms:
\begin{itemize}
    \item Dynamic cost indexing (DCI): modifying the speed of the flight to speed up and recover delay or, in some cases, even to slow down to save some fuel while maintaining the route \citep{dci}. Note that a change of cost index will imply not only a change on the cruising speed but also on the location of the top of descent, generally increasing the cruise and reducing the slower descent.
    \item Waiting for passengers (WfP): actively delaying outbound flights to wait for delayed connecting passengers. This option is currently seldom used by airlines, as it impacts the on-time performance of the outbound flights and, in some cases, waiting for passengers might lead to outbound flights being regulated. However, when the optimal solution to minimise the cost of delay is sought, then this might be a relevant strategy \citep{cass_II, COOK201638}. This could be particularly important for passengers who will need to be rebooked on next-day flights if they miss their connection, leading to significant costs for the airline. 
\end{itemize}

Two different levels of implementation are considered for the DCI and WfP rules: a baseline implementation which aims at capturing the most common current practices of airline operators, and an advanced implementation where decisions are made considering expected costs and the use of coupled sub-mechanisms.

\subsubsection{Baseline implementation}

The baseline implementation uses a rule of thumb to approximate current practice in the airline industry for DCI and WfP. The parameters of these mechanisms are calibrated with feedback received from a number of industry experts.

The cost index is determined before the take-off (\ie at push back) and it is fixed throughout the flight. The AOC considers the departure delay 
and the delay recovery strategy is performed as follows: 
(i) no delay will be recovered tactically if the estimated departure delay is smaller than 15 minutes; (ii) the decision of recovering delay is taken with probability linearly increasing with the value of estimated departure delay, when it is between 15 and 60 minutes (probability of recovery goes from $0.2$ at 15 minutes to $1$ at 60 minutes); (iii) the flight will always try to recover delay if the estimated departure delay is larger than 60 minutes.

Once the AOC has decided to try to recover delay, it will try to recover as much delay as possible (up to 5 minutes). The maximum recovered delay is however limited by the amount of extra fuel that would be required, being capped at 70\% of the total amount of additional fuel available. Moreover, to align the application of this rule with current practices, the flight never speeds up to the maximum possible speed; rather, the maximum selected speed is capped at 90\% of the maximum velocity. Finally, after applying all of the so far mentioned constraints, if the amount of delay that can be recovered is lower than 5 minutes, no recovery is performed, as experts highlighted that recoveries of delay lower than 5 minutes are seldom performed.

Waiting for passengers is performed 5 minutes before the aircraft is ready for push-back. At that moment, the AOC checks which passengers are not at the gate ready for boarding, and it estimates the time required for them to arrive to the gate. For this estimate, the AOC uses the most up-to-date information available: the actual or estimated time of arrival of the inbound connecting passengers' flights and their minimum connecting time (MCT). The MCT is pre-calculated for each airport and it varies as a function of the type of connection: domestic - domestic, domestic - international, etc. In the baseline implementation, the AOC decides to wait for any passenger with a flexible ticket whose at-gate time is estimated to be at most 15 minutes after the expected push back time of the outbound flight.

\subsubsection{Advanced implementation}

In the advanced implementation of the mechanism, the decisions on DCI and WfP are purely driven by the advanced estimation of the expected costs. Before departure, the mechanism couples the assessment of DCI and WfP decisions via a unified cost function. However, at top of climb (TOC), a second optimisation is performed and the delay recovery strategy is reassessed.

In the first instance, an estimation of the costs associated to waiting or not-waiting of each inbound passenger group which is not present at the gate 5 minutes prior push back ready is performed. The cost of waiting considers the extra delay required to allow the passengers to make their connection (considering the estimated arrival time and the MCT). The cost of not-waiting considers all types of passengers' cost due to missed connection (\eg duty of care, re-booking, soft costs). The required waiting times for all the passengers groups defines the range of departure delay that will be further analysed. For example, let us consider a flight which has an estimated departure delay of 20 minutes, with the maximum delay that can be recovered (selecting the maximum speed) being 8 minutes. Then, there are 2 groups of passengers which are late for boarding with estimated delays of 10 and 15 minutes, respectively, \ie waiting extra 10 minutes will allow for the boarding of the first group of passengers, while waiting extra 15 minutes will ensure that both groups can board the plane. In this situation, the AOC assess all possible costs associated with departing delays in the range of 35 minutes, together with the application of any possible delay recovery strategy: from departing with 35 minutes of delay, thus waiting for both passenger groups, but with no delay recovery, to departing with 20 minutes of delay and arriving with 12 minutes of delay, that is not waiting for passengers and recovering the maximum possible delay.

All the estimated costs are added and observed as recoverable delay (from not recovering any delay to the maximum number of minutes a flight can recover by speeding up and using the extra fuel available).
This is done by the use of dynamic cost indexing. DCI is considered to estimate the fuel cost required to achieve each of the different potential recovery times. Differently from the baseline implementation, the AOC might decide to recover only part of the delay and there is no limitation on the maximum velocity that the flight can choose in order to recover delay, as the decision is purely driven by the cost and the objective is finding the optimal solution given the estimated costs and physical constraints.

Combining the cost of waiting or not waiting for passengers, and the cost of recovering delay, a joined solution is provided. The decision that minimises the total cost is selected, \ie a decision to wait (or not) and to recover delay (or not) is performed before push back. This approach enables more complex decision such as waiting for passengers and then recovering part of the delay, trading a reduction on cost associated to missed connections with an extra cost on fuel consumption. 
When the flight reaches the TOC, a re-assessment of the expected arrival delay and potential delay recovery is performed. In this case, some of the cost associated with the departure delay will already have been accounted (\eg cost of re-accommodating passengers who missed their connections). However, some further costs could still be managed, in particular the ones related to the arrival delay. At this moment, the AOC considers the scheduled inbound time (SIBT) to estimate the arrival delay, in particular the potential arrival buffers are taken into consideration. Accordingly, the speed is selected to minimise the total cost, and the advanced 4DTA incorporates the possibility to slow down the flight at TOC in order to save fuel, when the expected arrival time is at least 30 minutes before the SIBT. This could be possible, in effect, due to the buffers, wind uncertainty, etc. This extra option enables the trading of unnecessarily long buffer for fuel savings.

\subsection{Flight prioritisation (FP)}
\label{subsec:FP_desc}
The flight prioritisation mechanism enables the swapping of ATFM slots at arrival airports regulations. This mechanism is inspired by UDPP principles \citep{UDPP_SID_2016}.

\subsubsection{Baseline implementation}
In the baseline implementation, ATFM slots are not allowed to be swapped.

\subsubsection{Advanced implementation}
In the advanced implementation, airlines are allowed to swap ATFM arrival slots between their own flights, but also exchanging their arrival slots with flights of another airline. When a flight plan is submitted to the Network Manager, if ATFM has been assigned due to a regulation at the arrival airport, the airline considers the possibility of the flight swapping. The following conditions need to be fulfilled to accomplish the swapping: (i) both flights have to be in the same regulation at the arrival airport, and (ii) the estimated cost of the swap needs to be negative (\ie the swap has a positive impact overall).

The cost of the swap is estimated as the total expected cost of delay if the slots are swapped minus the cost of delay if they are not, \ie cost${_1}$(slot${_2}$) + cost${_2}$(slot${_1}$) - (cost${_1}$(slot${_1}$) + cost${_2}$(slot${_2}$)), where slot$_{1}$ is ATFM arrival slot assigned originally to the first flight, slot$_{2}$ is the ATFM arrival slot for the second flight, and cost${_1}$ and cost${_2}$ are the delay cost function of the first and second flight for a given arrival slot. 

In order to estimate the cost of delay associated with each slot, the most up-to-date information is used, \eg considering the next available flight for missing connections. These costs consider passenger (hard and soft costs) and non-passenger related costs (crew and maintenance). Other network effects, such as an estimation of reactionary delay or the impact on curfew, are also considered. The model could estimate exactly which subsequent flight will be impacted by the propagation of delay, in terms of aircraft and/or connecting passengers, since all information is known to the airlines (or can be accessed by them). However, this is computationally too expensive increasing by a large factor the simulation time. As a consequence, a heuristic is used for the decision-making process of swapping the slots. This is done by computing the number of flights that will be operated by the same aircraft and assuming that a similar cost to the one experienced by the current flight will be incurred by them in a worst case scenario approach.

A novel contribution with respect to current operations comprises allowing flight swaps between different airlines, thus introducing the exchange of information between airlines for cost estimation. This is, in effect, an unrealistic assumption for the current real-world operational environment, where such information about costs is never shared by airlines. However, a market mechanism (\eg credit system, auction) can be devised, in principle, for an efficient slot exchange, with benefits for both the counterparties, for example by providing a payment to the airline delaying its own flight. Here, the presence of such market mechanism is assumed to justify the possibility of flight swapping between two airlines. However, a possible implementation is out of scope of this paper and it is left for future research.

\subsection{Flight arrival coordination (FAC)}
\label{subsec:FAC_desc}

The flight arrival coordination mechanism focuses on the sequencing done by E-AMAN systems at airports. This mechanism is only implemented in airports which have or are expected to operate an E-AMAN (24 airports as according to the SESAR Pilot Common Project \citep{PCP}). For these airports there are two moments when flights are issued delay when approaching them to manage their arrival and landing sequence:
\begin{enumerate}
    \item when the flight enters the planning horizon of the E-AMAN (with a default value of 200~NM from the airport),
    \item when the flight enters the execution or tactical horizon of the E-AMAN (defined at 120~NM from the airport).
\end{enumerate}
\noindent
The distances selected for the planning and execution horizon are in accordance with the expected extension of the arrival managers from 100-120~NM to 180-200~NM \citep{PCP}. 

When a flight enters the planning horizon, using an optimisation function which depends on the level of advancement of the mechanism, all the flights located in the scope of the arrival manager, \ie between the planning and the execution horizon, are re-assigned to available arrival slots, \ie not tactically assigned yet. The flight which triggers this optimisation, \ie the one which enters the arrival manager, receives the amount of delay that it is expected to require during its sequencing and approach. This flight then tries to absorb as much delay as possible by slowing down (saving some fuel). 

At the execution horizon, the flight which exits the E-AMAN triggers a similar re-optimisation and the final slot is assigned. The required delay (if any) is then performed as holding. 

Note that the optimisation problem is solved by considering all flights within the E-AMAN domain, each flight entering or leaving the system triggers a new optimisation which determines the optimal sequence. However, the new optimal solution may disrupt the previous sequence, thus successive optimisation problems might lead to some inefficiencies. Furthermore, in some cases, the slot assigned at the planning stage might not be available anymore when the flight reaches the execution horizon due to the stochastic dynamics of the system.

\subsubsection{Baseline implementation} \label{subsec:EAMAN_baseline}
In the baseline implementation, the flight arrival coordination tries to minimise the amount of holding delay that will be carried out at the TMA by minimising the total holding delay. The FAC optimisation function focuses on the maximisation of the arrival throughput at the runway. No information from the airlines is considered when optimising the arrival sequence. When a flight enters the planning horizon, the first slot available, considering the estimated landing time of the flight, is assigned. In a similar manner, once the flight enters the execution horizon, the first available slot is assigned and the holding delay computed.

\subsubsection{Advanced implementation}

In the advanced implementation, the FAC tries to minimise the expected total cost for each flight, considering the cost of fuel estimated by the flight (considering potential fuel savings achieved by slowing down, but also extra fuel cost by performing holding), and the cost of delay estimated by the AOC (for passenger and non-passenger related costs of delays).

\noindent
The FAC considers the expected cost of each slot for each flight. This information is provided to the FAC by the flight:
\begin{enumerate}
    \item when the flight enters the E-AMAN planning horizon, the FAC sends a list of slots available to the flight;
    \item then the flight requests the expected cost of delay for each slot to its AOC;
    \item in parallel, the flight estimates the fuel required for each slot (including potential savings by slowing down and required holding times) considering aircraft performance and operational parameters (\eg current weight);
    \item finally, the total expected cost is computed considering both, the cost of delay (from the AOC) and the cost of fuel (computed by the flight). These costs per available slot are then relied to the FAC to be used in the optimisation.
\end{enumerate}

\noindent
Once the FAC receives this expected costs per slot, the arrival sequence is optimised with the objective of minimising the total expected cost for all flights within the E-AMAN scope, and delay issued to the incoming flight. When the flight reaches the execution horizon, the same optimisation is performed with the already provided expected cost per slot.

\section{Indicators}
\label{sec:indicators}

Mercury is able to produce detailed low-level metrics. An analysis of the raw output of the model is required to gain an understanding of the effect of introducing the mechanisms in the system. This is achieved through the use of standard indicators, and also advanced network metrics.

\subsection{Standard indicators}
A set of classical metrics, variously used in ATM, intended to capture fundamental statistics regarding delays and costs (both for flights and passengers) are considered. Note that metrics which relate to passengers represent an advance with respect to classical flight-centric metrics. In fact, as shown in previous research \citep{POEM}, these passenger metrics provide a deeper understanding of the trade-offs involved in ATM operations. Examples of these metrics are: departure and arrival delay; number of delayed and cancelled flights; reactionary delay (number of flights and average magnitudes); and, passenger delay considering connecting and non-connecting passengers.

\subsection{Centrality}
In a networked system, such as ATM, centrality is a measure of the `importance' of a node in terms of its role in connecting the network. 
In the network of airports and flights, flights represent links and are therefore present only in some time intervals. 
By considering the difference between scheduled and actual (operated) flights, two air traffic networks can be defined. The effect of a mechanism to improve the robustness of the system (or parts thereof) 
can be assessed by analysing the differences between centrality in the scheduled and actual network, as the mechanism should mitigate disruptions in network connectivity. Most centrality metrics are not suited for these types of analysis \citep{Mazzarisi2018} as they focus on static, single-layer networks, while air traffic is more naturally described by a temporal, multilayer network \citep{zaninlillo}. Links in the network dynamically appear and disappear, as these links represent flights, and connections between them are only possible in the right temporal order. All the flights from the same alliance (or airline) can be considered as a different layer of the network \citep{Zaoli2019}. Both these aspects, the temporal and the multi-layer structures, must be considered to measure the disruption of network connectivity caused by delays. The failure of existing centrality metrics to capture these effects is analysed in detail in \citep{Mazzarisi2018}.
\\
Two tailored centrality metrics are proposed. Inspired by the Katz centrality \citep{Katz1953}, it is considered that the centrality of a node depends on the number of itineraries in the network having that node as origin (outgoing centrality) or as destination (incoming centrality). Only itineraries (or `walks') which are time-respecting and which consider the time required to travel through a link, are considered, \ie only itineraries which can be travelled (considering flight duration and transfer times) should be counted. Two different ways of weighting itineraries are proposed.

First, `trip centrality' (TC) \citep{Zaoli2019}: an itinerary of $n$ legs outgoing from (incoming to) an airport contributes $\alpha^n$ to the outgoing (incoming) centrality of that airport, where $\alpha<1$, so that longer itineraries contribute less. If the itinerary comprises flights of different airlines, its contribution is reduced by a factor $\varepsilon^m$, where $m$ is the number of changes of layer through the itineraries, and $\varepsilon<1$. This considers that multi-airline itineraries are less useful. 
In summary, the itineraries considered by trip centrality are all the possible itineraries that could be used by passengers (they are temporally feasible), weighted higher when they have fewer legs and fewer changes of airline.
    
Second, `passenger centrality' (PC): each itinerary contributes to the outgoing (or incoming) centrality of an airport: an amount which corresponds to the number of passengers on that itinerary. Therefore, the outgoing passenger centrality of an airport corresponds to the number of passengers that take a flight departing from that airport. The incoming centrality is defined by the number of passengers that arrive at the airport, either as final destination or with further connections. 

The damage to network connectivity due to disruptions can be estimated as the loss of centrality between the scheduled and the operated network. The operated network considers delays and cancellations, therefore an airport's trip centrality is always smaller than its centrality in the scheduled network. The loss of outgoing trip centrality of an airport therefore measures the loss of potential outgoing itineraries that become unfeasible in the operated network, quantifying the decrease in the potential to access the rest of the network from that airport (see \citep{Zaoli2019} for more details). 

For passenger centrality, in the operated network only passengers reaching their destination using their scheduled itinerary are considered. Therefore, passengers with a modified itinerary imply a loss of passenger centrality. 
If, for example, $N$ passengers incoming to airport $i$ miss their connection at $i$, and are rebooked to another flight, airport $i$ will have a loss of outgoing centrality amounting to $N$. The same loss would apply if $N$ passengers depart late from $i$ and miss their next connection at another airport. Therefore, the loss of outgoing passenger centrality of an airport accounts for the passengers that experience a disruption at that airport and for those that experience problems downstream. The actual incoming passenger centrality corresponds to the number of passengers that were scheduled and managed to reach the airport following their scheduled itineraries. Therefore, the loss of incoming passenger centrality can be interpreted as the impact to airport $i$ caused by issues upstream. 

\subsection{Causality}
Delays and congestion states propagate through the system due to the interactions between flights and the environment. Causality metrics aim to reveal the channels of delay propagation, identifying which nodes are facilitating the spreading process, for instance by forming subsystems working as amplifying feedback for delay propagation.

In statistics, detecting a (directional) causal relationship between two time series is equivalent to assessing whether the information on the past states of one variable helps in forecasting the future state of the other. Two causality metrics, Granger causality in mean \citep{granger1969investigating} and Granger causality in tail \citep{hong2009granger} are considered. 
The difference between the two metrics is that the latter considers only `extreme' states (large delays, in our application), therefore assessing whether there is a significant dependence between the extreme states of two variables. The former considers the dependence between all states.
Given two random variables $X$ and $Y$, whose states at different times are captured by two time series:
\begin{itemize}
\item $Y$ `Granger-causes in mean' $X$ if we reject, at some confidence level, the null hypothesis that the past values of $Y$ do not provide statistically significant information about future values of $X$, by assuming a linear VAR(p) process as the predictive model (see \citep{granger1969investigating} for further details on the implementation of the method);

\item $Y$ `Granger-causes in tail' $X$ if we reject, at some confidence level, the null hypothesis that the past {\it extreme} values of $Y$, defined as states falling in the (right) tail of the distribution\footnote{Thus, we can consider the time series of binary variables, which are $1$ if the state is extreme, $0$ otherwise.}, do not provide statistically significant information about future {\it extreme} states of $X$ (for further information, see \citep{hong2009granger}). 

\end{itemize}

We consider the network of airports and flights, where airports represent the nodes and a (directional) link is present if a causal relationship between two nodes exists. The state of delay of the airport is defined as the third quartile of the delay distribution of flights departing within a one-hour time window at each node.

Given $N$ airports, the network of causal relationships is built by applying the Granger causality test (`in mean' or `in tail') to all the possible $N(N-1)$ pairs of airports. We use the false discovery rate method \citep{benjamini1995controlling} to correct for false positives, due to the number of tests carried out \citep{tumminello2011statistically}, while maintaining an overall significance level equal to $5\%$. Similarly, the use of the Bonferroni correction has been considered in other publications (\cite{zanin2017network,Mazzarisi2018}).

Granger causality metrics thus measure how delays at an airport propagate to other airports. The number of causal links quantifies the global amount of delay propagation and the structure of the causality networks tells us how delays are propagated.
The topology of the network of causal relationships in the ATM system is paramount to the understanding of the dynamics of delay propagation at the macro level and to investigate whether the introduction of the mechanisms represents an improvement, \eg by disrupting some propagation channels, thus restoring optimal functioning. To this end, topological network metrics can be extracted from the network of causal relationships, \eg link density, reciprocity, clustering. In this research, we consider: (i) {\it link density}, \ie the number of causalities with respect to the number of all possible causalities, capturing the average level of causality in the process of delay propagation; (ii) {\it reciprocity}, measuring the likelihood of nodes being mutually linked, thus indicating subsystems of delay amplification;
and, (iii) the number of {\it feedback triplets}\footnote{A feedback triplet is a subgraph of three nodes, A, B, and C, where A causes B, B causes C, and C causes A.}, representing subsystems of three airports where delay propagates circularly.

The impact of implementing a mechanism at the network level can be assessed by measuring the changes in causality and disruption to feedback effects with respect to a baseline. However, the value of a network metric, such as the number of feedback triplets, depends on the link density of the network, which may change from one scenario to another. For a fair comparison, we consider the `over-expression' of such a metric, where over-expression is defined as the ratio of the observed value of a network metric to its expected value in the random case of an Erd\"os-Renyi graph with the same link density.

In evaluating the variation of such network metrics from the baseline to some scenario with a new ATM mechanism implemented, it is crucial that the adopted test of Granger causality does not display a high false positive rate, especially in relation to the considered subgraphs, \ie two-way linked nodes or feedback triangles. Recently, it has been proved that the test of Granger causality in tail by \citep{hong2009granger} is sensitive to autocorrelation of time series, thus resulting in a high false rejection rate of two-way causality in the presence of unidirectional relationships \citep{mazzarisi2020tail}. The authors of \citep{mazzarisi2020tail} introduced a novel test of Granger causality in tail, based on the bivariate version of the vector discrete autoregressive process for binary time series, referred to as `BiDAR', which solves this issue while maintaining very good power and adequate size. In the following, we use this novel test for the analysis of simulations of the agent-based model.

\section{Scenarios}
\label{sec:scenarios}

The entire ECAC space is modelled considering commercial flights on 12\textsuperscript{th} September 2014. This includes approximately 27k flights and 3.4M passengers (distinguishing between flexible and non-flexible ticket holders), with approximately 800 airports. This date and dataset was carefully selected to be representative of a high-traffic, non-disrupted day in 2014, with demand thus similar to an average day in 2023 (STATFOR baseline forecast). This allow us to extrapolate that the results presented would be valid to future operational days. The traffic is based on historical DDR data, schedules and generated passenger itineraries, 
and calibrated using historical data from CODA and DDR \citep{coda,EUROCONTROL_DDR2} as described in Section~\ref{subsec:calibration}.

Two scenarios (`hub delay management' and `effect of E-AMAN scope') are designed to evaluate the mechanisms.

\subsection{Hub delay management}

This scenario has been designed to evaluate the impact of the 4D Trajectory Adjustment and the flight prioritisation mechanisms to alleviate the impact of disruptions at main hubs. As indicated in Section~\ref{subsec:4DTA_desc}, 4DTA is composed of two sub-mechanisms (dynamic cost indexing and waiting for passengers), even if these mechanisms could be applied by any airline operating at any airport, their maximum benefit is directly related to the expected cost (and benefit) of managing disruptions driven by passenger connectivity. Airlines without connecting passengers will only benefit from the use of dynamic cost indexing, while airlines with passenger connections tend to concentrate them at their hubs. For these reasons,  this scenario considers that a large disruption is impacting three large hubs in Europe: Amsterdam (HEAM), Heathrow (EGLL), and Paris Charles de Gaulle (LFPG). The disruptions are modelled by manually defining an ATFM regulation at each airport with the following characteristics: regulations start and finish in the morning (06:00 -- 14:00 local time), and the airport capacity during the regulated period is half of the average nominal capacity of the airports (45 arrivals/hour for HEAM, 54 arrivals/hour for EGLL and 44 arrivals/hour for LFPG).

In addition to the imposed ATFM regulation, the rest of the delay in the system is set as default (\ie nominal day of operations) and ATFM regulations are defined at other airports based on a randomly selected nominal day.
    
This scenario is executed in: (i) a baseline form (where all three mechanisms are implemented in baseline), and (ii) using 4DTA and FP with the advanced implementation.

Note that even if the manually set disruptions are only defined for the three hubs, the mechanisms are implemented in the whole system, \eg if the 4DTA is implemented in the advanced mode, this mechanism will be exploited by all flights regardless of their airline type, and origin and destination.

\subsection{Effect of E-AMAN scope} 
\label{subsec:scenario_scope}
The second case study focuses on the different implementations of the flight arrival coordination mechanism. The FAC is implemented in the baseline and the advanced version, the latter considering different planning horizon ranges, in order to analyse the role of the extension. With larger horizon, all flights are considered earlier in determining the landing sequence: on one hand, an early decision may lead potentially to more fuel savings, but, on the other hand, it tends to reduce flexibility in combination with higher uncertainty.

Hence, in order to analyse the role of the horizon range, the network is simulated with high average delays. In particular, the delays assigned by the ATFM regulation are modelled as random selection from the distributions characterising historical days of operations with high number of assigned ATFM delays, reduced capacities of the airports and longer en-route and taxiing operations. 

In total four different cases are executed (two baseline and two advanced FAC implementations):
\begin{itemize}
    \item Baseline with nominal and extended range: the FAC is implemented in the baseline version as described in Section~\ref{subsec:EAMAN_baseline}, considering both 200~NM and 600~NM for the planning horizon.
    \item Advanced FAC with nominal and extended range: the advanced version of the FAC is implemented, with the planning horizon equal to 200~NM or 600~NM.
\end{itemize}

\section{Results}
\label{sec:results}

In this section, we show the results of the comparison between the simulations of the ATM system with the novel mechanisms implemented and the baseline version of the model as described in Section~\ref{sec:scenarios}. Most of the analyses are performed considering 100 simulations of the model. For the centrality analyses, 50 simulations are used. The results of each metric are averaged across these, and in the following discussion, these averages are presented. The baseline scenario is the reference point, for both classical and network metrics we show their percentage change with respect to the baseline. As a robustness check, subsamples of 50 independent simulations are considered and their results compared: thus unless differently specified, the results shown are those that are consistent in the subsamples. Note that the results presented are for all flights modelled.

\subsection{4D Trajectory Adjustment -- Hub delay management}
\begin{figure}
    \centering
    \includegraphics[width=1\textwidth]{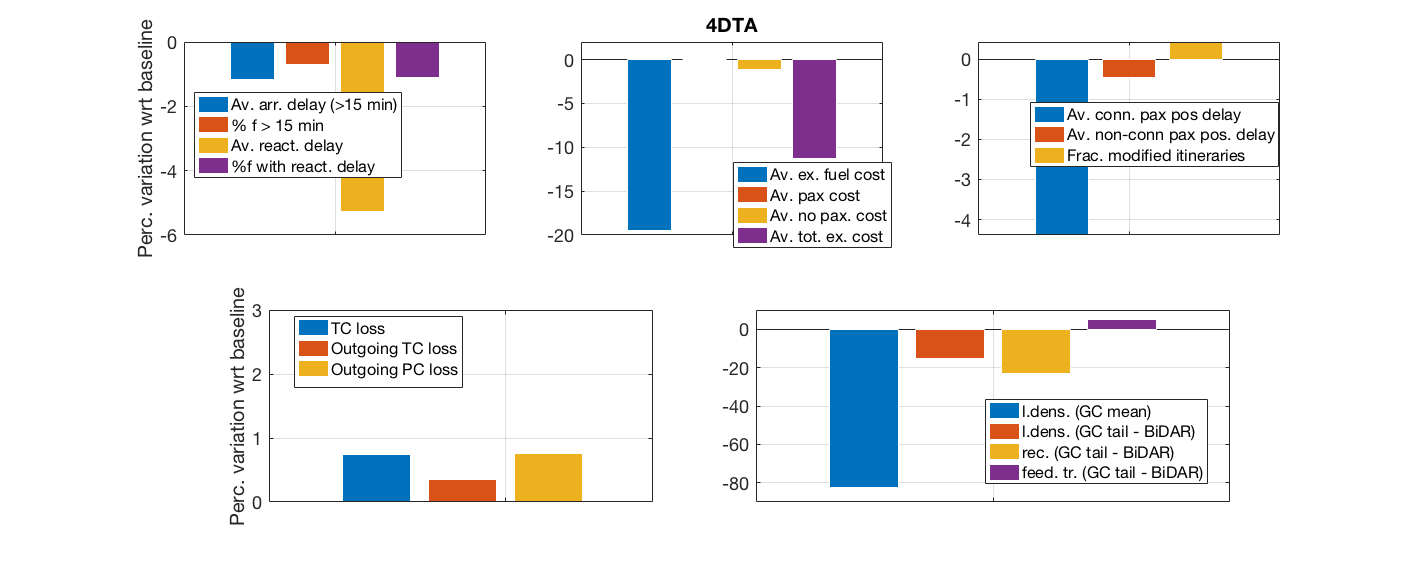}
    \caption{Summary of percentage changes of metrics in 4DTA with respect to the baseline. Panels show, respectively, metrics related to delay, cost, passengers, centrality and causality. Abbreviations used in legends: Av. = average, arr. = arrival, $\%$f > x min = percentage of flights with delay larger than x minutes, react. = reactionary, ex. = excess, pax = passenger, conn. = connecting, pos. delay = positive delay (early arrival counted as zero), TC = trip centrality, PC = passenger centrality, l.dens = link density, GC = Granger Causality, rec. = reciprocity, feed.tr. = feedback triplets. }
    \label{fig:4DTA_summary}
\end{figure}
\begin{figure}
    \centering
    \includegraphics[width=1\textwidth]{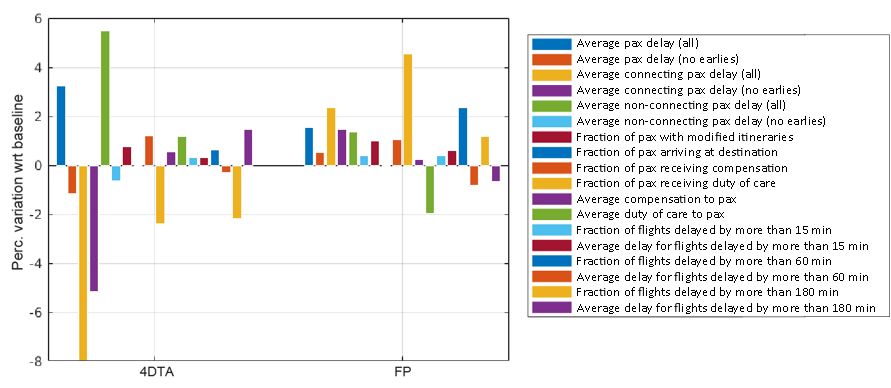}
    \caption{Percentage change in passengers-related metrics advanced 4DTA with respect to the baseline.}
    \label{fig:Hub_perc_pax}
\end{figure}
\begin{figure}
    \centering
    \includegraphics[width=0.5\textwidth]{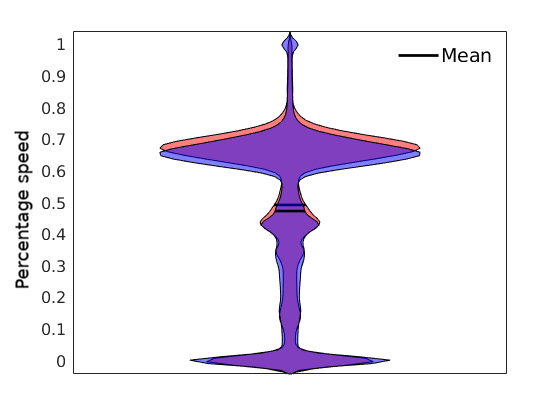}
    \caption{Violin plot of the percentage speed selected when applying 4DTA at TOC by flights with at least one connecting passenger (orange) and by flights not carrying any connecting passenger (purple).}
    \label{fig:4DTA_perc_speed}
\end{figure}

Figure~\ref{fig:4DTA_summary} shows the percentage change of a selection of classical and network metrics when 4DTA is implemented in its advanced form in the `hub delay management' scenario with respect to the baseline. \\
The top-left panel shows that the introduction of the 4DTA mechanism improves the system by reducing the flight delays. This is true for all the displayed quantities, namely the average arrival delay for flights with more than 15 minutes of delay, their fraction, and the reactionary delay (number of flights and amount of delay). This is consistent across different measures of delay (arrival and departure delays $\geq X$ with $X \in \{0,15,60,180\}$, not shown). This suggests that the possibility to adjust speed is useful to partially absorb large delays. \\
The top-centre panel shows that, with 4DTA, there is a sizeable reduction of excess fuel cost (up to almost $20\%$). This can be understood by the fact that flights use 4DTA to control in a more efficient way their total costs, and the cost of fuel is a significant factor driving the solution. Such behaviour creates a cost difference with respect to the baseline implementation, where the decision to recover delay does not consider the economic impact but only the delay magnitude. Other types of cost are only marginally affected by the introduction of 4DTA, \eg the average passenger-related costs, in some cases even increasing. Non-passenger delay costs (\eg cost of maintenance and crew associated with delay) display a small but sizeable decrease. Overall, the costs are reduced by more than $10\%$ when 4DTA is introduced in the system.\\ Positive passenger delays are also reduced (see top-right panel). This is much more evident when considering connecting passengers (blue bar) rather than non-connecting ones (red bar), since they benefit more by the introduction of 4DTA. In fact, the departure delays created by the regulations are only recovered when it is advantageous cost-wise (\eg if there are connecting passengers missing their connections due to the delay). Although the fraction of modified itineraries (\ie passengers which are re-booked due to missed connections) slightly increases, passengers are better off when 4DTA is in operation. More passenger metrics are shown in Figure~\ref{fig:Hub_perc_pax}. We note that the average passenger delay (including early arrivals, counted as negative) increases with respect to baseline. This is due to having less early arrivals, since 4DTA allows flights to slow down to reduce fuel costs, trading buffers and early arrivals for fuel savings. The same effect can be seen on the average non-connecting passenger delay. We also note that the reduction in positive delays is driven by a reduction in the fraction of flights with delays $\ge 180$ min. A reduction of the fraction of passengers receiving duty of care and a smaller increase of the average duty of care produce an overall improvement.\\
The improvements for connecting passengers can be due to a more efficient use of the waiting for passengers option or to a speeding up of the flights with connecting passengers on board. Regarding the latter factor, we verified that the flights carrying at least one connecting passenger select slightly higher percentage speeds than flights that contain no connecting passengers when applying 4DTA to speed up (see Figure~\ref{fig:4DTA_perc_speed}). We also note that the use of waiting for passengers is quite different in the baseline and advanced implementation of the mechanism. In fact, although a similar number of flights considers applying the waiting for passengers option in the two scenarios, only 15$\%$ of them actually apply it in the advanced scenario, against 27$\%$ in the baseline, and the average the wait is shorter (3.8 minutes against 6.7 minutes in the baseline). This suggests that the strict rule used in the baseline scenario (\ie always waiting for any flex passengers if the wait is below 15 minutes) is not efficient cost-wise when compared to the more flexible strategy of the advanced scenario. In fact, flex passengers (7.7$\%$ of the 3.4M passengers) are widely spread over flights (67.9$\%$ of flights have at least one flex passenger) and often they are connecting passengers (21.6$\%$ of them), and this leads to a large use of waiting for passengers in the baseline. This should be adjusted as part of a future calibration. The strategy used in the advanced scenario, evaluating the cost of waiting for passengers, leads to a more moderate use of the option, which however seems more efficient both for costs (as seen in the previous paragraph) and for connecting passengers.

The causality and centrality metrics (bottom panels of Figure~\ref{fig:4DTA_summary}) partly confirm the view given by the classical metrics. The centrality metrics display a slightly larger loss, possibly in connection with the larger number of modified itineraries. In fact, modified itineraries are due to disrupted connections, which impact both trip centrality and passenger centrality losses.  Causality metrics signal that the level of distress propagation between airports is lower with 4DTA. In fact, the density of causal links shows a very significant decrease according to both types of causalities, in mean and in tail. This suggests that delays in one airport less frequently cause delays in another, both for average and extreme delays, because the 4DTA mechanism tends to absorb most of the delays, especially for connections, leading to  softening of the propagation. A further confirmation of such improvement is the reduction of network reciprocity: the two-airports feedback effects, measured by reciprocity, also decrease substantially, meaning that a propagated delay less frequently propagates back, creating an amplification effect. There is, however, a small increase in the feedback triplets. This overall improvement in delay propagation is certainly linked to the decreased delays, especially reactionary delays, as well as to a more refined waiting for passengers mechanism. In fact, the large use of waiting for passengers in the baseline is a potential source of delay propagation, where the delay of an incoming flight causes the delay of an outgoing one.

In summary, the introduction of the 4DTA mechanism makes the system better off from the point of view of airlines, passengers and the environment (due to reduced fuel consumption). The system is more efficient (from the cost and delay perspective) and more robust to local shocks at airports, which propagate much less. These results are aligned with previous findings in the literature \citep{cass_II,COOK201638}.

\subsection{Flight prioritisation -- Hub delay management}
\begin{figure}
    \centering
    \includegraphics[width=1\textwidth]{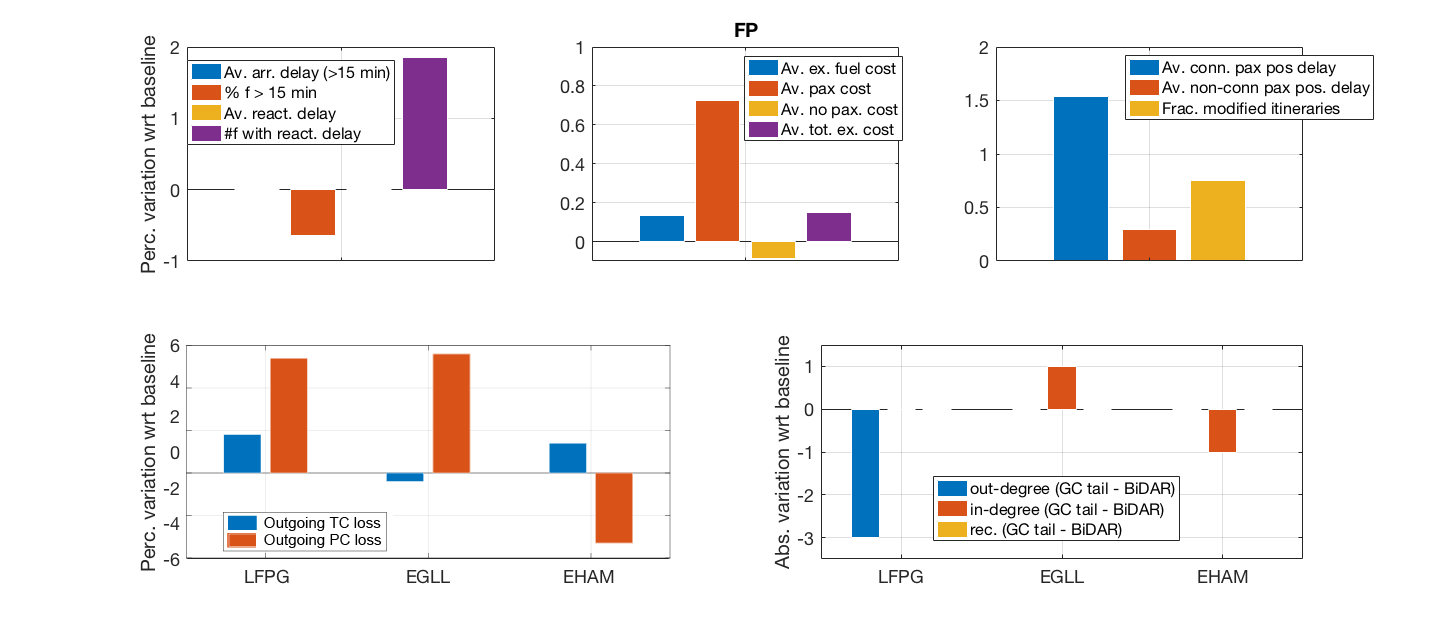}
    \caption{Summary of percentage changes of metrics in FP with respect to the baseline. Panels show, respectively, metrics related to delay, cost, passengers, centrality and causality. The percentage variations of the classical delay and cost metrics are restricted to the three airports where the FP mechanisms are applied and we consider all flights landing at any of the three airports within the time window of active regulation.  Abbreviations used in legends: Av. = average, arr. = arrival, $\%$f > x min = percentage of flights with delay larger than x minutes, react. = reactionary,  $\#$f = number of flights, ex. = excess, pax = passenger, conn. = connecting, pos. delay = positive delay (early arrival counted as zero), TC = trip centrality, PC = passenger centrality, GC = Granger Causality, rec. = reciprocity.}
    \label{fig:FP_summary}
\end{figure}

The `hub delay management' scenario has also been tested with the introduction of the flight prioritisation mechanism. Figure~\ref{fig:FP_summary} shows the percentage change of a selection of classical and network metrics in this scenario with respect to the baseline. Classical metrics (top panels) indicate that the system is worse off overall. There are more reactionary delays (top left), total costs are higher mainly due to fuel and passenger costs (top centre), passenger delays and modified itineraries are slightly increased (top right). It is important to note, however, that these variations are quite small (never larger than $2\%$, often much smaller). This suggests that the introduction of FP has little or no impact, at least when measured with classical metrics. As FP is implemented across the network in the simulation but can only be used when flights are affected by ATFM regulations at arrivals, we also measured the effects restricted to the three hubs which were manually disrupted. However, this means restricting the analysis to a small number of flights, and the results were not statistically significant, in that they were not consistent on two independent sets of 50 model iterations. Therefore, no statistically significant effect are detectable in the airports where the FP mechanism is implemented.

Considering the entire system, centrality metrics show increased losses, in line with the larger fraction of modified itineraries. Causality metrics, despite some changes of the node degree of the airports, \ie the number of causal links pointing to or outgoing from the node, display no significant variations of reciprocity and feedback triplets, which supports the hypothesis that FP has negligible effect on delay amplification, thus it is less effective than 4DTA at the systemic level. The bottom panels of Figure~\ref{fig:FP_summary} show centrality and causality metrics restricted to the three disrupted hubs. The sign of variations is not consistent across the airports, nor between two sets of 50 independent model iterations, therefore again we cannot detect any significant effect when restricting to these airports.

In summary, the introduction of FP appears to have essentially little or no effect (or maybe slightly negative) at the system level, when considering delay, cost, centrality, and causality. More surprisingly, at the level of the three manually disrupted hubs, no significant effect can be detected, suggesting that the effects, if any, are very small and essentially stochastic fluctuations. 

\subsection{Flight arrival coordination -- Effect of E-AMAN scope}

As presented in Section~\ref{subsec:scenario_scope}, two different sub-scenarios are evaluated: one with E-AMAN operating with a nominal radius of scope of the planning horizon (\ie planning horizon located at 200~NM from the airport), and one with an the extended radius (\ie planning horizon located at 600~NM).

\begin{figure}
    \centering
    \includegraphics[width=1\textwidth]{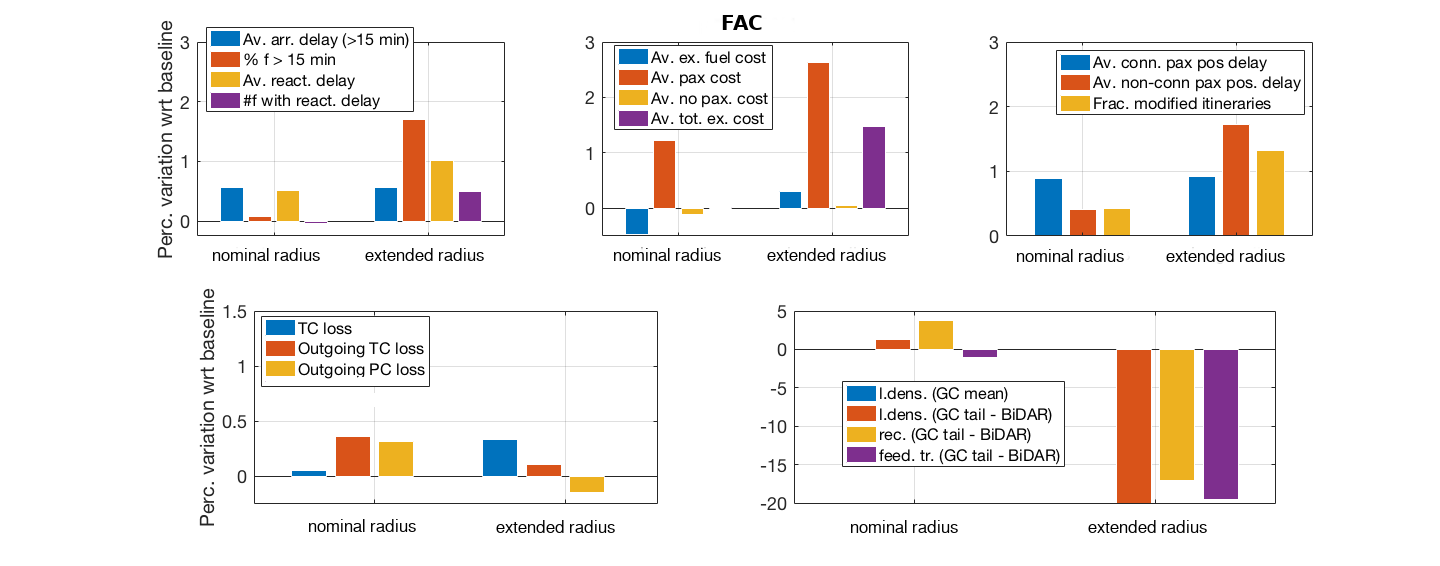}
    \caption{Summary of percentage changes of metrics in FAC with respect to the baseline. Panels show, respectively, metrics related to delay, cost, passengers, centrality and causality. The percentage variations of both classical and centrality metrics are restricted to the airports where the FAC mechanisms are applied. Only flights landing at any airport of the restricted sample in computing the delay and cost metrics are presented. The causality metrics are extracted from the network of causal relationships between any couple of the ECAC airports, but considering only the subgraphs, \ie the reciprocated links or the feedback triangles, involving at least one airport with the FAC mechanism.  Abbreviations used in legends: Av. = average, arr. = arrival, $\%$f > x min = percentage of flights with delay larger than x minutes, react. = reactionary, $\#$f = number of flights,  ex. = excess, pax = passenger, conn. = connecting, pos. delay = positive delay (early arrival counted as zero), frac. = fraction,  TC = trip centrality, PC = passenger centrality, l.dens = link density, GC = Granger Causality, rec. = reciprocity, feed.tr. = feedback triplets.}
    \label{fig:EAMAN_summary}
\end{figure}

The results presented are restricted to the analysis of the 24 airports where an E-AMAN is operating  (see Section~\ref{subsec:FAC_desc}). In particular, we consider only flights landing at an airport among these 24 in computing delay and cost metrics. Figure~\ref{fig:EAMAN_summary} shows the percentage variations of a selection of metrics for the two choices of radius with respect to the baseline where the network is simulated in a congested operational day. The advanced implementation of the FAC increases the delays of both flights (top-left panel) and passengers (top-right panel). The implementation with the nominal range increases the average arrival delay and reactionary delay. The extended radius setting appears to further worsen the delay situation, increasing also the number of flights with delay and reactionary delay. As a consequence, passenger costs are larger (top-centre panel), especially in the extended radius setting, producing higher total costs. Quite surprisingly, the excess fuel cost is only very slightly smaller (in the nominal radius FAC setting) or even larger (in the extended radius setting) than in the baseline scenario. Thus, it seems that the introduction of the FAC mechanism in its advanced form makes the system less efficient.

\begin{figure}
 \subfloat[Planned extra fuel (fuel savings) expected at planning horizon for flights issued with delay.\label{fig:EAMAN_planned}]{%
   \includegraphics[width=0.45\textwidth]{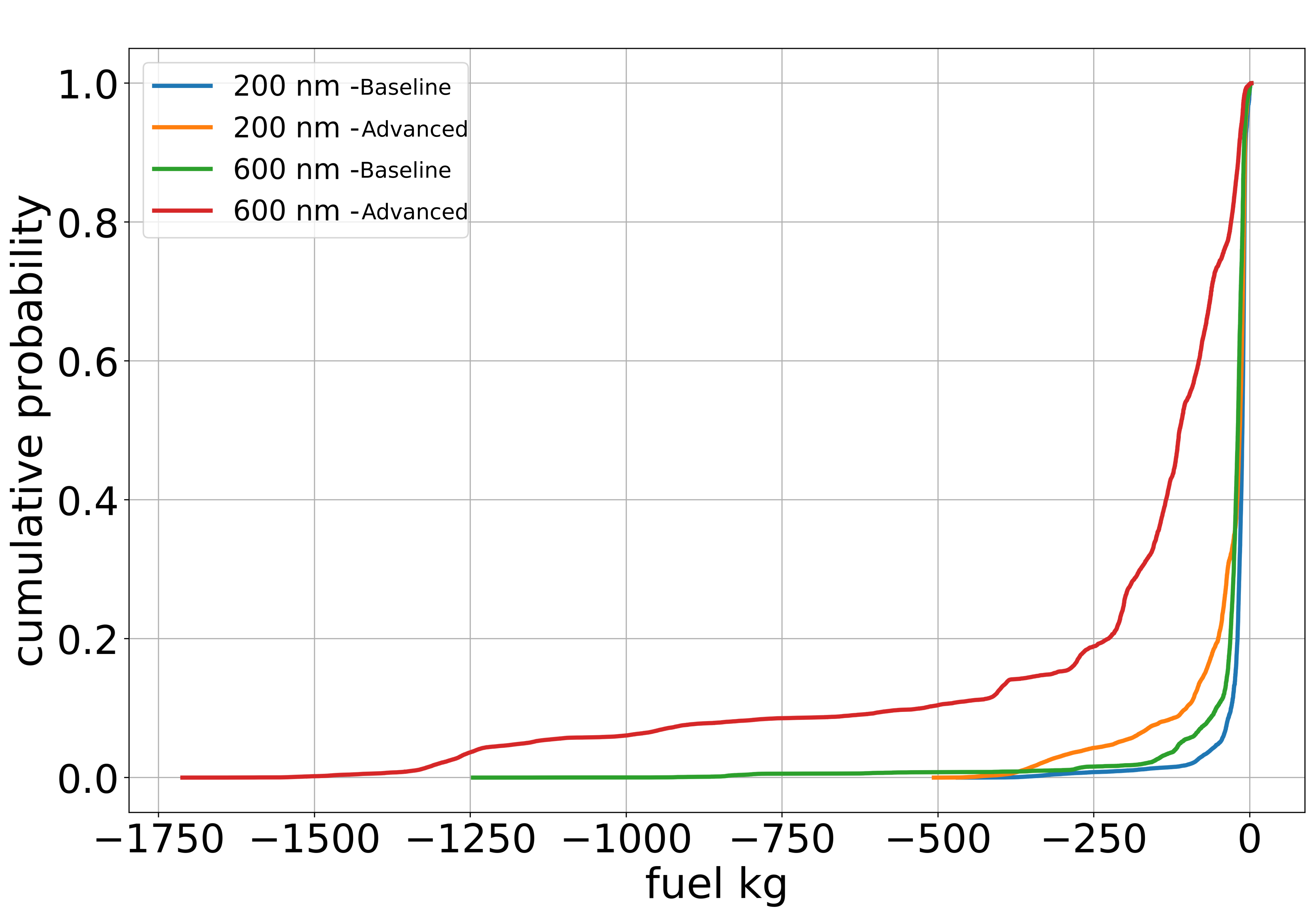}
 }
 \hfill
 \subfloat[Extra delay required between delay assigned at planning and execution horizons.\label{fig:EAMAN_extra_tact}]{%
   \includegraphics[width=0.45\textwidth]{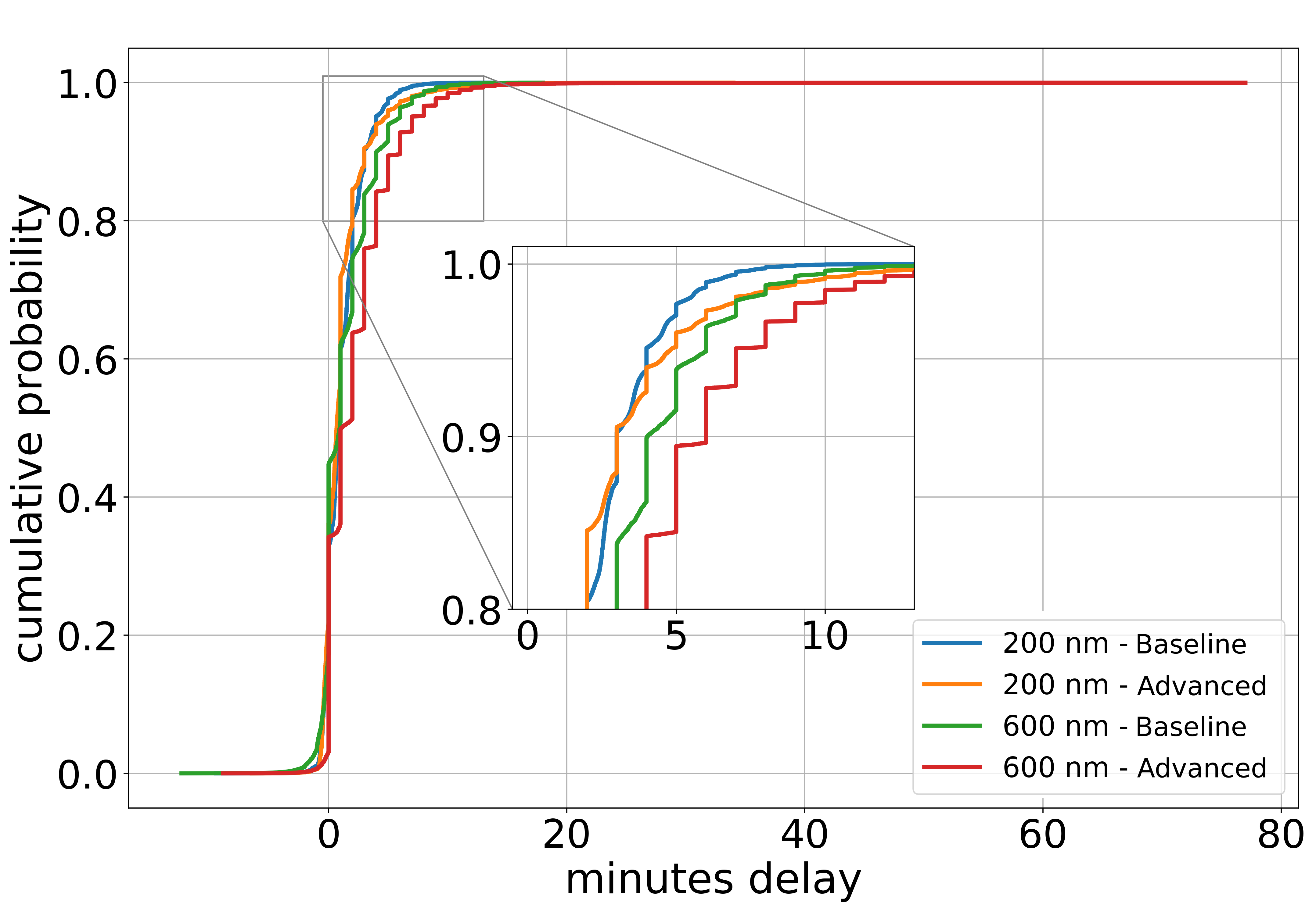}
 }
 \caption{Metrics at planning and execution horizon in E-AMAN}
 \label{fig:dummy}
\end{figure}

This can be attributed to a discrepancy between the E-AMAN planned and actual holding time, which causes the assignments of additional holding delays to respect the planned landing sequence. In fact, the advanced implementation of E-AMAN assigns, in some cases, large delays (\eg delays of over 20 minutes). This happens, as shown in Figure~\ref{fig:EAMAN_planned}, partially because the advanced FAC is trying to minimise the expected total cost, and in some cases, some delay might represent fuel savings as the cruise speed is reduced. This is particularly relevant for large aircraft on the larger extended scope scenario. In the baseline implementation, as the radius increases from 200 to 600~NM, the fuel that is potentially saved increases: expected total fuel saved by slowing down to avoid holding with a 200~NM scope in the baseline implementation is on average $19.6$ $t$ ($2$ $kg$ per flight), with 600~NM radius, these expected saving increase to $24.2$ $t$ ($2.5$ $kg$ per flight). In the advanced implementation, the expected savings are even larger ($32.7$ $t$ (average of $3$ $kg$ per flight) at 200~NM, and $269.0$ $t$ (average of $27$ $kg$ per flight) at 600~NM). As the advanced implementation is trying to minimise the cost, later slots might be assigned (even if at the expenses of delay) to save fuel. \\
However, these expected benefits are not observed in the final results, as presented in Figure~\ref{fig:EAMAN_summary}. The planned landing sequence is broken due to not considering forthcoming demand and particularly due to the lack of managing of uncertainties. Each time a flight enters one of the radii of the E-AMAN, the landing sequence is optimised. However, there is no capacity reserved for new arrivals which are not already in the E-AMAN scope; and more importantly the flight might experience non-accounted delays due to uncertainties while in the E-AMAN meaning that it might not be able to use its planned slot. The consequences are discrepancies between the delay issued in the planning horizon and the actual delay required when the flight reaches the execution horizon as shown in Figure~\ref{fig:EAMAN_extra_tact}, \ie when the flight arrives to the execution horizon it might be delayed due to uncertainty and not able to use its planned slot, or its planned slot might not be available as assigned to other flights, leading to holdings.\\
This effect is particularly relevant for large aircraft frames, as they are more prone to save fuel by slowing down their cruise, and therefore to receive larger delay at the planning horizon. However, these flights will also have higher holding fuel consumption and more passengers on-board, so an overall worsening of fuel and passengers indicators is observed when further delay is assigned to them at the execution horizon if, for example, they miss their planned slot due to tactical uncertainties.

In summary, the naive approach to enhance the E-AMAN by re-optimising the landing sequence, considering expected costs, produces an overall worsening with respect to focusing on minimising the arrival delay, as uncertainties and future demand are not properly considered.

This is further confirmed by the network metrics. Centrality metrics (bottom-left panel of Figure~\ref{fig:EAMAN_summary}) show small and positive variations, meaning that the introduction of the mechanism makes the centrality loss of these airports larger. This is likely due to the increase of modified itineraries and more generally to the increased delays. The causality metrics (bottom-right panel of Figure~\ref{fig:EAMAN_summary}) are extracted from the network of causal relationships between all the couples of the ECAC airports, but considering only the subgraphs involving at least one airport where the mechanism is implemented. Here the outcome is more difficult to interpret: while the introduction of the advanced FAC with the nominal radius E-AMAN makes the system slightly worse off, in the extended radius scenario the system becomes significantly less connected, from a causal point of view, both in terms of the number of causal links and of feedback effects (reciprocated links and triplets). This could be explained by the fact that FAC increases the arrival delay of flights independently of their departure delays, thus masking the causal relationships due to network effects.

In conclusion, the introduction of the advanced FAC mechanism appears to make the system worse off from the point of view of airlines and passengers, as well as from a systemic perspective as a whole. With the exception of causality, all the metrics are worse for the extended range, than for the nominal range. Furthermore, this conclusion holds even when considering the whole ECAC space, and not only the 24 airports where the mechanism is implemented.

\section{Conclusions and further work}
\label{sec:conclusions}

Various advanced mechanisms inspired by operational concepts, with a special emphasis on new rules based on an airline's network-wide cost minimisation, have been explored in this paper. The Mercury agent-based model allows us to run a simulation of a single day of operations, tracking passengers and aircraft, modelling the complex interactions and decision making processes of the stakeholders, and, in particular, airlines. The micro-level nature of the model supports the quantification of various metrics related to different stakeholders, and assessing overall network performance under various scenarios with the use of standard and advanced network (centrality and causality) metrics. 

The `hub delay management' case study showed that the advanced 4D trajectory adjustments (4DTA) mechanism, adjusting aircraft departure times and cruise speeds, is a powerful tool in the airline's operations. 4DTA is an efficient solution to reduce costs for the airline, as well as the average delay for flights and passengers. Connecting passengers benefit greatly from this, while non-connecting passengers see their positive arrival delay only slightly decrease and their arrival delay (including early arrivals) increase. Hence, there is a trade-off between airline economic efficiency and (some) passenger utility, but, on the whole, passengers are better off with 4DTA. The existence of such trade-offs is important to highlight before deployment, and should nurture the debate on implementation priorities. Centrality tends to worsen at the network level with 4D trajectory adjustments, however, probably due to a higher number of modified itineraries. Propagation of delay between airports, as measured by causality metrics, tends to decrease with the mechanism implemented. That is, local disruptions affect fewer other parts of the network. The 4DTA mechanism has proven much more efficient at the network level than the advanced flight prioritisation (FP), which has only small and slightly negative effects overall, and no statistically significant effect for the airports at which the mechanism is implemented.

The analysis of the `effect of E-AMAN scope' case study has highlighted how an advanced implementation of the flight arrival coordination (FAC) requires the consideration of uncertainties and forthcoming demand. The advanced, but naive, optimisation implemented, where the sequence of arrival slots is re-optimised considering expected costs each time a new flight enters the arrival manager scope has proven inefficient. Flight are assigned to later slots at the planning horizon to save fuel by slowing down but uncertainties lead to missed slots and extra holding delay.   
Thus leading to a non-optimal behaviour of the system where the saved fuel is then used as holding and delay at arrival is also experienced. This effect is aggravated as the scope of the E-AMAN is increased as uncertainty increases.

The main high-level goal of the Domino project was to provide a tool and a methodology to analyse the interdependencies between the elements of the air transportation system. This paper presented how this was achieved by the development and use of Mercury, and the computation of network-wide advanced metrics of centrality and causality. The metrics developed and evaluated in Domino could be further used at different levels, for instance analysing the coupling of subsystems in the network. Further work is also required to gain a deeper understanding on the relationship between these new metrics used in this paper and established ones. These network metrics may be good proxies for others, and/or add further dimensionality and usefulness, but this can only be understood via statistical analyses and additional case studies. 

\section*{Acknowledgment}
This work has been performed as part of the Domino project which has received funding from the SESAR Joint Undertaking under grant agreement No 783206 under European Union’s Horizon 2020 research and innovation programme. The opinions expressed herein reflect the authors’ views only. Under no circumstances shall the SESAR Joint Undertaking be responsible for any use that may be made of the information contained herein.

\bibliographystyle{apalike}

\bibliography{mybib.bib}{}

\end{document}